\documentstyle[aps,preprint]{revtex}   
\begin{document} 
\draft 
\title{Low energy properties of $M$-state tunneling 
systems in   metals: New candidates for non-Fermi-liquid 
systems} 
\author{G.  Zar\'and}  
\address{Condensed Matter Physics Research Group of the 
Hungarian Academy of Sciences \\
at the Institute of Physics, Technical University of   
Budapest,  
P.O.\ Box 112, H-1525, Hungary.} 
\date{\today}  
\maketitle 
\begin{abstract}  
We construct a generalized  multiplicative 
renormalization group transformation  to study the 
low energy dynamics of a heavy particle tunneling among 
$M$ 
different  positions and interacting with $N_f$ 
independent conduction electron channels.  Using a 
$1/N_f$-expansion we show that this M-level scales towards a 
fixed point equivalent to the $N_f$ 
channel 
$SU(M) \times SU( N_f ) $ Coqblin-Schrieffer model. 
Solving numerically the scaling equations we find that a 
realistic M-level system  scales close to this  
fixed point (FP) and its Kondo temperature  
is in the experimentally 
observable range $1-10 K$. 
\end{abstract} 
 

One of the simplest examples of non-Fermi-liquid (NFL)
systems is provided by the two-channel Kondo model
\cite{Noz_Bland,Ludwig_Affl,Andrei}, where the two
conduction electron channels overcompensate the spin (or
some quasi-spin) of an impurity and a new degenerate
ground state appears with an intermediate coupling
\cite{Noz_Bland,Ludwig_Affl,NRG}.  The FP
corresponding to the ground state is characterized by
non-zero residual entropy \cite{Andrei} and universal
power low behavior in the impurity resistivity and the
energy dependence of the scattering rate ($R_{\rm
imp}\sim T^{1/2}$, $1/\tau \sim \omega^{1/2}$)
\cite{Ludwig_Affl}. Several normal
state properties of $Ce$ or $Y$-based heavy fermion
compounds like $CeSn_3$ and $YbCuAl$ can also be
explained surprisingly well in terms of the 2-channel
Kondo effect \cite{Cox_hfsup}.

Other extensively studied possible realizations of the 
two-channel
Kondo model are provided by fast two-level systems
(TLS's) in metals \cite{Kondo,Vlad_Zaw}.  These TLS's are
formed by some heavy particles (HP's) tunneling between
two neighboring sites and interacting strongly with the
conduction electrons.  Recent experimental
\cite{Ralph_Ludwig} and theoretical
\cite{Hershfield,Zar_Vlad} investigations confirm the
conjecture that at low temperatures these TLS's can be
properly described by an effective two-channel Kondo
model, where the spatial motion of the HP is coupled to
the angular momentum of the conduction electrons via an
effective exchange coupling and the two degenerate
channels correspond to the two real spin directions of
the conduction electrons
\cite{Ralph_Ludwig,Hershfield,Zar_Vlad}.

The TLS model sketched above is appropriate to describe
tunneling centers in amorphous metals \cite{Phillips},
%
however, it breaks down in systems like the narrow gap
semiconductor $Pb_{1-x}Ge_xTe$ or  insulating
$K_{1-x}Li_{x}Cl$ alloys \cite{Fukuyama,MLS} where the
tunneling centers are formed by some substitutional
impurities, and the HP is tunneling between 3, 6 or 8
equivalent positions.   Therefore the question arizes in a
natural way, what is the low temperature behavior of an
M-level system (MLS) which strongly interacts with the
conduction electrons.

To answer this question we consider a model previously 
introduced to describe the effect of the excited states 
in a 
TLS \cite{Zar_Zaw}.  In the following we assume that the 
temperature (or the relevant energy scale) is low enough 
and 
thus the motion of the HP is restricted to the lowest 
lying 
$M$ states corresponding to the $M$ spatial positions of 
the 
HP: 
\begin{equation} 
H_{hp} = \sum_{i,j=1}^M b^+_i \Delta^{ij} b_j \; , 
\end{equation}  
where $b^+_i$ creates a HP at site $i$ and $\Delta^{ij}$ 
is 
the tunneling amplitude between positions $i$ and $j$.  
We 
assume that no external stress is present and that the 
$M$ 
positions are completely equivalent thus $\Delta^{ii}=0$. 
 
The most general two-particle interaction generated by 
the 
screened Coulomb interaction (or a pseudopotential) 
between the HP and the 
conduction electrons takes the form: 
\begin{equation} 
H_{\rm int}= \sum_{\scriptstyle i,j,n,m  
\atop\scriptstyle 
\epsilon, \epsilon^\prime} b^+_i V_{nm}^{ij} b_j 
a^+_{\epsilon   n   f} a_{\epsilon^\prime   m   f}  
\; , 
\end{equation}  
where $a^+_{\epsilon n f}$ creates a conduction electron
with energy $\epsilon$, orbital quantum number $n$, and
'flavor' index $f$ ($f=1,..,N_f$).  The orbital quantum
number $n$ can be thought of as the angular momenta of 
the
electrons while for a real MLS the quantum numbers 
$f=1,2$
correspond to the spin up and spin down conduction 
electrons
and $N_f=2$.  For the sake of simplicity we also assume a
constant density of states between the high and low 
energy
cutoffs $D$ and $-D$, $\varrho_0$, for all flavor 
numbers.
Naturally, both the couplings $V^{ij}_{nm}$ and
the tunneling amplitudes $\Delta^{ij}$ are connected by 
the
symmetry properties of the MLS which will be exploited 
later
on.

Following similar lines as in Ref.~\cite{Abrikosov} one 
can 
introduce the HP propagator ${\cal G}^{ij}(\omega)$ and 
the 
HP-conduction electron vertex function 
$\Gamma^{ij}(\omega)$ 
in the usual way.  However, calculating these functions 
in a 
perturbative way it turns out that they do not satisfy 
the 
usual multiplicative renormalization group (RG) 
equations. 
Therefore we consider the following generalized 
RG transformation ($T=0$): 
\begin{eqnarray} {\cal G}(\omega, v^\prime,  
\Delta^\prime, 
D^\prime) & = & A \; {\cal 
G}(\omega, v,\Delta, D)\; A^+ \; , \nonumber \\  
\Gamma(\omega, v^\prime, \Delta^\prime, D^\prime) & = &  
[ A^+]^{-1}  
\Gamma(\omega, v,\Delta,D) \; A^{-1} \; , 
\label{eq:mrg}  
\end{eqnarray}  
where the matrix notations $\varrho_0 
V^{ij}_{mn}\rightarrow 
v$, $ \Gamma^{ij}_{mn}\rightarrow \Gamma$, and 
$\Delta^{ij} 
\rightarrow \Delta $ have been introduced, $D^\prime$ 
stands 
for the scaled bandwidth and $A$ is an $M\times M$ matrix 
acting in the HP indices.  Note that  $A=A(v^\prime, 
\Delta^\prime,  
D^\prime/D)$ is 
independent from the dynamical variable $\omega$. 
 While for finite $D/D^\prime$ the matrix $A$ has a rather 
complicated structure for an infinitesimal change of $D$ it
can be chosen to be Hermitian and Eq.~(\ref{eq:mrg}) can 
be cast in the form of a scaling equation for the 
dimensionless couplings $v^{ij}$. 
 
In the following we assume that the relevant energy variable 
is $\omega$, i.e., $\omega\gg |\Delta^{ij}|, T$.  In this 
case the inverse HP propagator and the vertex functions can  
be expressed in the next to leading logarithmic order  
\cite{Vlad_Zaw} as 
\begin{eqnarray} 
({\cal G}^{-1})^{ij} & = & \omega \; \delta^{ij} -  
\Delta^{ij} +  N_f \ln {D\over \omega} \;   
(\delta^{ij}\;\omega \;{\rm tr}\{ v^{kl} v^{lk} \} - {\rm  
tr}\{ v^{ik} \Delta^{kl}  
v^{lj} \}) \nonumber \\ 
\varrho_0 \Gamma^{ij} & = & v^{ij} - \ln{D \over \omega}  
\;  
\Big( [v^{ik},v^{kj}] - N_f {\rm tr} \{ v^{ik}  v^{lj}  
\}v^{kl}  
\Big) \; , 
\label{eq:verex} 
\end{eqnarray} 
where $[\phantom{m},\phantom{m}]$ denotes the commutator, 
the trace operator ${\rm tr}\{...\}$ is acting in the 
electronic indices, and a summation must be carried out 
over repeated indices. 
 Plugging (\ref{eq:verex}) into Eq.~(\ref{eq:mrg}) one can 
easily generate the scaling equations in a selfconsistent  
way: 
\begin{eqnarray} 
{d\Delta^{ij}\over dx} & = & -{1\over 2} N_f \Big[ {\rm  
tr}\{  
v^{ik} v^{kl} \} \Delta^{lj} + \Delta^{ik} 
{\rm tr}\{ v^{kl} v^{lj} \}  - 2 {\rm tr}\{ v^{ik}  
\Delta^{kl}  
v^{lj} \}  
\Big] \; \nonumber \\ 
{dv^{ij}\over dx} & = & - [v^{ik},v^{kj}] + {1 \over 2}  
N_f  
\Big(2 {\rm tr} \{ v^{ik}  v^{lj} \}v^{kl} - {\rm tr}\{  
v^{ik}  
v^{kl} \}  
v^{lj} - v^{ik} {\rm tr}\{ v^{kl} v^{lj} \} \Big)\;, 
\label{eq:scaling} 
\end{eqnarray}  
where the scaling variable $x=\ln{(D_0/D)}$ has been 
introduced.  Since the scaling equations are rather complicated apart 
from some special cases they  
can be solved only numerically. However, to exploit the  
symmetry properties of the MLS it is useful first to  
introduce some site representation in the orbital  
indices of the conduction electrons. This can be  
achieved most simply by taking some linear combinations  
of the most strongly scattered angular momentum channels  
and hybridize them by using group theoretical methods. 
Working only with electron states 
directed to  the impurity positions the $v^{ij}$'s become  
$M^4$-dimensional tensors. However, the number of  
independent couplings is largely reduced by  symmetry.  

 A typical scaling of the norm of the dimensionless  
couplings, 
$u=\sum ||v^{ij}||$, is shown in Fig.~\ref{fig:U's} 
(dashed 
line).  The initial couplings have been estimated by 
using 
similar methods as in Refs.~\cite{Vlad_Zaw,Zar_Zaw}. 
Similarly to the multichannel Kondo problem both the 
infinite and the weak coupling FP's are unstable and 
the system scales to an intermediate strong coupling 
FP \cite{Noz_Bland,Ludwig_Affl,NRG}.  The Kondo energy 
can be identified as the crossover energy from the weak to 
strong coupling regimes:  $T_K = D_0 \; e^{-x_c}$, $x_c$ 
being the crossover value of the scaling parameter.  For 
realistic initial parameters we find that this Kondo 
temperature can easily be found in the experimentally 
observable region, $T_K \sim \; 1- 10\;K$. 
 
In order to determine the properties of the MLS in the  
regime $T,\omega \ll T_K$ one has to identify the FP the MLS 
scales to \cite{remark}. In the following we shall show that 
a MLS scales towards a FP, which -- up to  
some potential scattering part -- has the same structure  
as the $SU(M)\times SU(N_f)$ Coqblin-Scrieffer model  
\cite{Coqblin}.  To prove this we first remark that  the operators  
$\delta^{ij}\sum_k v^{kk}_{nm}$ and $\delta_{mn}\sum_p 
v^{ij}_{pp}$ are  
invariant under scaling. Therefore $v^{ij}$ can be  
written as $v^{ij}_{mn} = {\tilde v}^{ij}_{mn} + 
M^{ij}_{mn}$, where  
the matrix $M$ is a constant of motion depending on the  
initial parameters and $\sum_i {\tilde v}^{ii}_{nm} =  
\sum_n {\tilde v}^{ij}_{nn} =0$. Then one can easily show  
that the right-hand side of Eq.~(\ref{eq:scaling})  
disappears provided 
\begin{equation} 
{\tilde v}_0^{ij} = {1\over N_f} \left(\matrix{O^{ij} & 0 \cr 0 & 
0 \cr}\right)\; , 
\label{eq:fp} 
\end{equation} 
where the $O^{ij}$'s are unitary equivalent to the  
generators of the $SU(M)$ Lie algebra, 
\begin{equation} 
[O^{ij},O^{kl}] = \delta^{il} O^{kj} - \delta^{kj}  
O^{il}\;. 
\label{eq:Lie}
\end{equation} 
with $O^{ij}_{nm} \sim  
\delta^i_m \delta^j_n - {1\over M} \delta^{ij}  
\delta_{nm}$.
Similarly to the TLS problem beside the one in Eq.~(\ref{eq:fp})
Eq.~(\ref{eq:scaling}) has an infinite number of FP's associated 
with  different 
reducible and irreducible representations of the $SU(M)$ 
Lie-algebra 
(\ref{eq:Lie}). Our numerical simulations show, however, that all 
the FP's which 
correspond to a representation  different from the defining one 
(\ref{eq:Lie}) are unstable.
In Fig.~\ref{fig:U's} we show the scaling of the 'algebra  
coefficient' 
$\alpha= \sum_{i,j,k,l} || N_f^2 [{\tilde v}^{ij},{\tilde 
v}^{kl}] - N_f  
\delta^{il} {\tilde v}^{kj} + N_f \delta^{kj} {\tilde  
v}^{il} || $, 
which measures in a natural way how well the fixed point  
(\ref{eq:fp}) is approached by the ${\tilde v}$'s.  
For $T<T_K$ (i.e. for  
$x>x_c$) the coefficient $\alpha$ scales to zero. Thus  
 {\it below the Kondo temperature an MLS  scales to the 
$SU(M)\times  
SU(N_f)$ Coqblin-Schrieffer model, which is a  
non-Fermi-liquid  model} and has a different scaling behavior 
than 
the 
2-channel Kondo model \cite{Ludwig_Affl,Ruckenstein}.
(For a 6-state system, e.g., we expect an $\omega^{1/4}$ 
behavior of the scattering rate.)

To show that FP (\ref{eq:fp}) is stable and to analyze its 
operator 
content
we write the deviations of the couplings from their FP value in a 
form
\begin{equation}
\delta v^{ij} =
\left(\begin{array}{cc}
\rho^{ij} & t^{ij} \\
(t^{ji})^+ & \lambda^{ij}
\end{array}\right)
\;, \label{eq:deviation}
\end{equation}
and linearize the scaling equations with respect to $\delta 
v^{ij}$. 
The 
couplings $\varrho^{ij}$, $t^{ij}$, and $\mu^{ij}$ are $M\times 
M$, 
$M\times 
\infty$, and $\infty \times \infty$ matrices, respectively.
Like the TLS case the linearized equations for $\varrho^{ij}$, 
$t^{ij}$, and 
$\mu^{ij}$ decouple completely,
\begin{eqnarray}
{d \mu^{il}\over dx} &=& {1\over N_f} \left( \delta^{ij}\mu^{kk} 
- 
M \mu^{il}\right)\;,
\label{eq:mscaling} \\ 
 {d\rho^{il}\over dx} & =& - {1 \over N_f} 
\left([O^{ik},\varrho^{kl}] + 
[\varrho^{ik},O^{kl}] \right) 
+ {1\over 2N_f} \Bigl\{2\delta^{il}\varrho^{kk} +  2 O^{jk}  {\rm 
tr} \{\varrho^{ij}  
O^{kl} + O^{ij}\varrho^{kl}\} 
\nonumber \\ 
&-& 2n\varrho^{il}
-O^{ij}{\rm tr}\{ \varrho^{jk}O^{kl} + O^{jk} \varrho^{kl}\} - 
{\rm 
tr}\{\varrho^{ij}O^{jk}
+O^{ij}\varrho^{jk}\}O^{kl}\Bigr\}\label{eq:roscaling} \\
{dt^{il}\over dx} &=& -{1 \over N_f} 
\left(O^{ik}t^{kl} - O^{kl}t^{ik}\right) + {1\over N_f}
\left(\delta^{il}t^{kk} -  M t^{il}\right)
 \;,\label{eq:tscaling}
\end{eqnarray}
and they can be solved exactly due to the simple structure of the 
$O^{ij}$'s.
These linearized equations  have an infinite number of zero 
modes; a 
finite number 
of them correspond to potential scattering while the others can 
be 
identified 
with the generators of the unitary transformations connecting the 
different 
possible $M$-dimensional subspaces where the $SU(M)$ Lie-algebra 
is 
realized.
These 0-modes can be shown, of course, to leave the Lie-algebra 
(\ref{eq:Lie}) 
unaffected. All the other modes can be shown to be irrelevant, 
thus 
the FP 
({\ref{eq:fp}) is stable.

The low energy properties of the MLS are determined by the 
operator 
content of 
the FP which is much richer and quite different from 
that of the simple Coqblin-Schrieffer model. A thorough analysis 
of 
Eq.~(\ref{eq:tscaling}) shows that for $M\ge 3$ the leading 
irrelevant operators
can be written as 
\begin{equation}
{\cal O}_l \sim \left(\begin{array}{cc}
0 & C^{ij} \\
(C^{ji})^+ & 0
\end{array}\right)\;,
\label{eq:leading}
\end{equation}
where  the $C^{ij}$'s  satisfy 
$\sum_l(C^{kl}_{mn}-C^{ml}_{kn})=0$. These operators scale
like $\sim T^{\lambda_l}$ with $\lambda_l=(M-1)/N_f$, and they 
describe scattering 
between channels which are not taken into account in the usual 
Coqblin-Schrieffer 
model. While they dominate the thermodynamical quantities like 
the 
specific heat, 
e.g., which scales  as $c_{\rm imp}\sim T^{2\lambda_l}$ they do 
not 
contribute to 
the resistivity, which scales like $\sim T^{\lambda_{sl}}$
with $\lambda_{sl}=M/N_f$, and is determined by subleading 
operators 
of the form
\begin{equation}{\cal O}_{sl} \sim \left(\begin{array}{cc}
Q^{ij}& 0 \\
0 & S^{ij}
\end{array}\right)\;,
\end{equation}
where the matrices $Q^{ij}$ and $S^{ij}$ satisfy $\sum 
Q^{ii}=\sum 
S^{ii}=0$
and $Q^{ij}_{mn} =Q^{ij}_{nm}$. Note that the operators 
(\ref{eq:leading}) do not exist in the TLS case ($M=2$), 
and therefore the low-energy properties of a TLS can be completely 
described by the two-channel Kondo model \cite{Zar_Vlad}.

At this point we have  to remark that since the FP couplings 
${\tilde v}^{ij}$ scale like $1/N_f$ the obtained scaling 
exponents  can be considered as the first order estimates in a 
$1/N_f$ expansion, and they become exact in the $N_f\to \infty$ 
limit \cite{Zar_Vlad,Gan_etal} ($\lambda_{sl} = M/N_f$ is 
e.g. the $1/N_f$-expanded version of the exact exponent 
${\tilde \lambda}_{sl} = M/(N_f + M)$ for $N_f>  
M$ \cite{Ludwig_Affl,Ruckenstein}).    However, 
the $1/N_f$ expansion breaks down for $N_f < M$, and 
while  for a physical MLS with $N_f=2$  we expect similarly to 
other  models \cite{Vlad_Zaw,Zar_Vlad} that the {\it fixed point  
structure}  remains the same,  it remains an open 
question  whether the new leading irrelevant operators 
(\ref{eq:leading}) 
survive in that  physical limit or  not. 
 
In order to decide whether the non-Fermi-liquid properties
can be recovered in reality or not  it is important to
study the scaling of the splitting $\Delta^{ij}$.  As soon
as the temperature (frequency) becomes smaller than the
renormalized splitting $|\Delta^{ij}(T,\omega)|$ the
dynamics of the MLS is frozen out and the Kondo effect
described above is stopped.  Therefore in order to observe
a non-Fermi-liquid behavior characteristic to the FP
(\ref{eq:fp}) we need $|\Delta^{ij}(T_K)| \ll T_K$.  The
scaling of the splitting parameter $\Delta^{12}$ is shown
in Fig.~\ref{fig:delta} (for the definition of
$\Delta^{12}$ see the inset in Fig.~\ref{fig:delta}).  As
one can see, for realistic model parameters
$\Delta^{12}(T_K) / \Delta^{12}(D_0)$ is as small as $\sim
10^{-3}$, and therefore, even for very large splittings
$\Delta^{12}\sim 100 K$ the splitting is strongly reduced
and since $\Delta^{12}(T_K)\ll T_K$ the MLS can get easily
in the vicinity of the 2-channel Coqblin-Schrieffer FP.

As recently pointed out by Moustakas and Fisher for TLS's 
\cite{Moust} multi-electron scattering becomes 
also relevant in the neighborhood of the two-channel Kondo fixed 
point. This process is also relevant in the present model in the 
vicinity of the NFL fixed point, however, similarly to the TLS 
case \cite{Zaw_etal} it has a small initial amplitude, and being 
irrelevant in the weak coupling region this amplitude is even 
more significantly reduced during the first part of the scaling, 
$T_K<D<D_0$. Therefore, in the neighborhood of the FP 
(\ref{eq:fp}) it is always the splitting $\Delta^{ij}$ discussed 
above which provides the dominant mechanism to drive the system 
away the NFL FP and the multi-electron processes play a less 
important role.

Concerning the experimental realizations there exist
already some experiments on non-commutative 8-state
systems in $Pb_{1-x}Ge_xTe$ alloys where the tunneling
centers are formed by the relatively small $Ge^{++}$ ions
\cite{Fukuyama}.  While in this material a Kondo
crossover has indeed been observed no non-Fermi-liquid
behavior has been detected.  However, in the experiments
the inter-impurity interaction was quite strong which
leads to the non-vanishing of the diagonal part
$\Delta^{ii}$ of the tunneling matrix.  Since, similarly
to TLS's, the differences $\Delta^{ii}-\Delta^{jj}$ are
renormalized much less then the offdiagonal matrix
elements \cite{Vlad_Zaw}, they can lead to the freezing
out of some impurity states corresponding to transitions
from the $SU(M)$ to $SU(M^\prime)$ models with
$M^\prime<M$, finally, most probably, reaching the trivial 
$M^\prime=1$ 
model.  Furthermore, in this material there is a
strong spin-orbit scattering, which leads to a strong
cross scattering between the spin up and spin down
electron channels and drives the system to the $N_f=1$
Fermi-liquid FP instead of the $N_f=2$
non-Fermi-liquid model.  Thus, for this material even if the 
inter-impurity
interactions were small we would expect a Fermi liquid
behavior.  Therefore it would be very interesting from
the experimental side to find some similar alloys with
{\it small spin-orbit interaction}, where the spin symmetry
would guarantee the non-Fermi-liquid properties. 
To  observe the non-Fermi liquid FP 
one could apply a pressure on the sample, which is an appropriate 
tool for tuning $T_K$  into the region 1-10$K$ \cite{Fukuyama}.

The author would like to acknowledge useful discussions  
with N. Andrei, K. Vlad\'ar and A. Zawadowski, and he  
would like to thank  for its hospitality the Institut  
Laue-Langevin, Grenoble,   where part of the present work  
has been done. This research has been supported by the  
Hungarian Grants OTKA F016604 and OTKA 7283/93.

\begin{figure} 
\caption{\label{fig:U's} Scaling of the norm of the  
dimensionless  
couplings, $u=\sum ||v^{ij}||$ (dashed line), and of the  
algebra  
coefficient $\alpha$ (continuous line) for a 6-state 
system  
with $N_f=2$.  
The initial couplings have been chosen to be  
$v^{11}_{11}=0.8$, $v^{11}_{22} = 0.2$,  
$v^{11}_{66}=0.1$, $v^{11}_{12}=0.05$,  
$v^{11}_{16}=0.03$, $v^{12}_{21} =v^{12}_{12} = 0.0005$,  
$v^{12}_{11} = 0.005$, $v^{16}_{11} = 0.003$,  
$v^{16}_{61} =v^{16}_{16} = 0.0005$. The other non-zero  
couplings have been generated by symmetry  
transformations.} 
\end{figure} 
 
\begin{figure} 
\caption{\label{fig:delta} Scaling of the  dimensionless  
hopping  
amplitude, $\Delta^{12}/D_0$ for the same 6-state system  
as in Fig.~1. Inset: Numbering of the  sites of the  
6-state system.} 
\end{figure} 
 

\begin{references}
\bibitem{remark} We assume in this part of the analysis that
$T$ or $\omega$ is larger than the renprmalized splitting.
\end{references}
\end{document}